# A 39pJ/b 7.3Gbps 1.3mm$^2$ Multi-Subcarrier Massive MU-MIMO-OFDM Detector Exploiting Beamspace Sparsity and Frequency-Domain Correlation in 22FDX

Abhishek Kumar†, Seyed Hadi Mirfarshbafan†, Oscar Castañeda, Christoph Studer

ETH Zurich, Zurich, Switzerland; †equal contribution

### Abstract

We present the first multi-subcarrier massive multi-user (MU) multiple-input multiple-output (MIMO) orthogonal frequency-division multiplexing (OFDM) data detector reported in the open literature. By exploiting the channel's beamspace sparsity and frequency-domain (FD) correlation, we achieve up to 3× area and power reduction. Our design supports $U = 8$ user equipments and $B = 64$ basestation antennas, computes soft outputs for QPSK to 256-QAM, and processes 16 subcarriers in parallel. The fabricated 22FDX ASIC has a core cell area of 1.3mm$^2$, consumes 286mW, and delivers a throughput of 7.3Gbps at 0.8V core voltage, achieving best-in-class energy efficiency of 39pJ/b.

### Introduction

Massive MU-MIMO and millimeter wave (mmWave) communication are key physical-layer technologies for 5G and upcoming wireless systems. Data detection in such systems entails high complexity, due to supporting a large number of antennas and wide bandwidths. OFDM—which is adopted in 5G and specified in 6G—further complicates data detection, as thousands of subcarriers must be processed either in parallel, which leads to excessive area and power, or serially, which incurs high latency and buffering requirements. We propose, to our knowledge, the first ASIC performing soft-output massive MU-MIMO detection for 16 subcarriers in parallel with best-in-class energy efficiency, facilitating replication to process larger number of subcarriers in parallel. To achieve high efficiency, we leverage two techniques as illustrated in Fig. 1. The first technique exploits the fact that channel impulse responses are concentrated in a few dominant taps, resulting in subcarrier correlation; this enables us to compute the linear minimum mean-squared error (LMMSE) equalization matrix only on a small subset of subcarriers, called *basepoints*, separated by $S$ subcarriers called *stride.* The equalization matrices for the remaining subcarriers are obtained via linear interpolation, enabling significant area and power savings. The second technique exploits the sparsity of mmWave channels (e.g., FR2 and FR3 in 5G) in the *beamspace* domain (BD) [1], which is obtained by applying a spatial fast Fourier transform (FFT) on antenna-domain (AD) received signals. In BD, both the channel matrix and the received signal are sparse, with energy concentrated in a few dominant beams. We leverage this sparsity to reduce the effective number of operations, resulting in power savings. BD signals, however, require larger bitwidths in their fixed-point (FXP) representation compared to AD signals. To mitigate this problem, we employ the variable-point (VP) number format [2].

### Architecture and Operation

The fabricated multi-subcarrier data detector ASIC consists of $N_p = 16$ parallel soft-output detector cores, six of which are illustrated in Fig. 2. Each detector core contains five main modules: The beamspace FFT performs AD to BD transformation on the OFDM demodulated input vector from all antennas. On each basepoint subcarrier $w \in \{1,6,11,16\}$, the basepoint processing unit (BPU) computes the LMMSE equalization matrix $\boldsymbol{W}$ and the terms required for log-likelihood ratio (LLR) soft-outputs, namely the inverse bias term $\mu^{-1}$, and the signal-to-interference-plus-noise (SINR) term $\nu$. The interpolation processing unit (IPU) computes $\boldsymbol{W}$, $\mu^{-1}$, and $\nu$ for non-basepoint subcarriers, through linear interpolation. The equalizer (EQ) calculates the symbol estimates, which are then used by the LLR unit to compute soft outputs for each bit using the max-log approximation.

Fig. 3 illustrates the BPU architecture, which contains a triangular systolic array of $U \times (U+1)/2$ processing elements (PEs) to compute the equalization matrix $\boldsymbol{W}$ in four phases: Gram computation, Cholesky decomposition, forward substitution, and backward substitution. For the systolic array to execute all four phases using the same PEs—whose architectures are shown in Fig. 4—the PEs contain multiple ports and multiplexers orchestrated by the controller.

Fig. 5 depicts the EQ architecture, consisting of $U$ complex multiply-accumulate (CMAC) units, each equipped with a sparsity-exploiting power saving mechanism, highlighted in red. A CMAC is skipped for certain inputs by disabling the registers whenever the magnitudes of both inputs are below the corresponding thresholds, resulting in dynamic power savings. Similar power saving circuitry is also built into the PEs (cf. Fig. 4). The bias engine inside the BPU uses an architecture similar to that of the EQ. The inputs to EQ and bias engine are converted to VP [2] via FXP2VP units to reduce the bitwidths of multiplier operands and converted back to FXP before addition, enabling 8% area savings compared to a fully FXP design. The architectures of the IPU and LLR unit are shown in Fig. 6. The IPU interpolates $\boldsymbol{W}$ column-by-column during each equalization cycle to avoid storing the full matrix. The timing schedule of the operations in Fig. 7 reveals that, as the number of equalization tasks increases, the relative overhead of preprocessing (BPU) diminishes, and system throughput is primarily determined by EQ.

### Implementation Results and Comparison

Fig. 8 depicts the die photo with the main components highlighted. The ASIC supports two modes: (i) AD mode with the beamspace FFT turned off and (ii) BD mode exploiting the beamspace sparsity to save power. In Fig. 8, we also show the coded block error rate (BLER) of the chip's output in both AD and BD modes with realistic line-of-sight (LoS) and non-LoS channels. As a baseline, we also show the BLER of a floating-point brute-force (BF) method in which all subcarriers are basepoints. Linear interpolation slightly improves BLER over BF due to implicit channel estimation denoising. Our ASIC's outputs achieve nearly the same BLER as the BF baseline, except in the challenging non-LoS 256-QAM scenario, with only a 1dB loss at the target 10% BLER specified in 5G.

Fig. 9 shows the measured power in AD and BD modes, and the estimated BF baseline power with both LoS and non-LoS stimuli. Compared to the BF baseline, interpolation enables 2.1× power and 3× area reduction (cf. Fig. 8). Exploiting beamspace sparsity enables up to 1.4× power reduction, yielding a total of 3× power reduction. The measurement results in Fig. 10 demonstrate that the chip achieves a max. clock frequency of 755MHz at the nominal voltage of 0.8V, corresponding to 7.3Gbps with 256-QAM (cf. Fig. 7). Forward body biasing boosts the max. frequency to over 900MHz, exceeding 8.8Gbps, at the cost of increased leakage power.

Table I compares the key metrics of our design with state-of-the-art MIMO detector ASICs. In addition to being the only design that processes multiple OFDM subcarriers in parallel, our design achieves best-in-class throughput and energy efficiency among designs that include preprocessing. Furthermore, the designs [4,5] implement low-complexity algorithms which suffer severe BLER degradation under realistic, correlated channels, as reported in the respective references and in [3]. By leveraging beamspace sparsity and FD correlation, our work presents a scalable multi-subcarrier detector solution for the massive MU-MIMO OFDM uplink.

**Acknowledgments** The authors thank GlobalFoundries for providing silicon fabrication through the 22FDX university program.

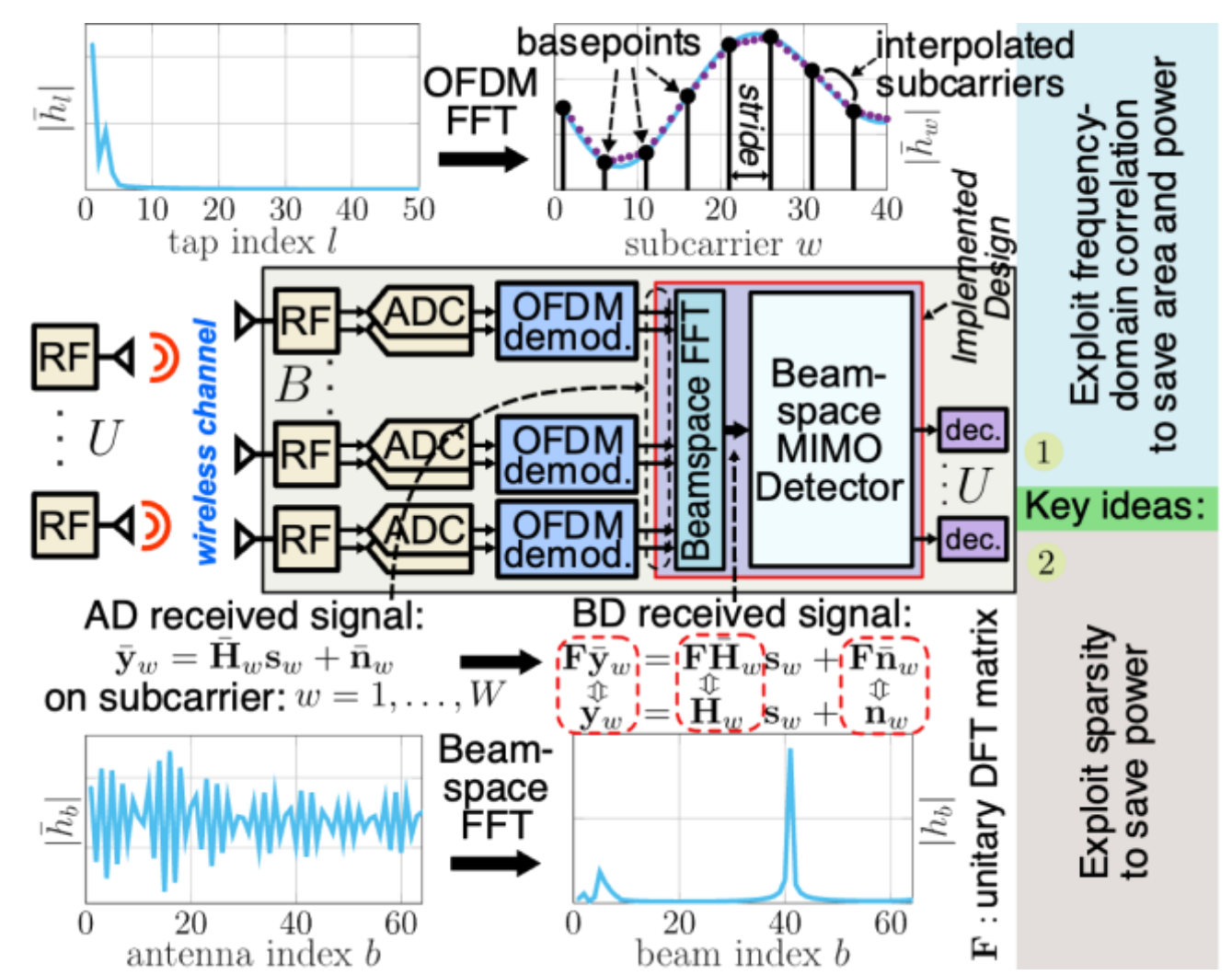

Fig. 1. Illustration of the massive MU-MIMO-OFDM uplink and the implemented design (middle); and the key techniques for area and power savings: (i) exploiting channel frequency-domain correlation (top); and (ii) beamspace sparsity (bottom).

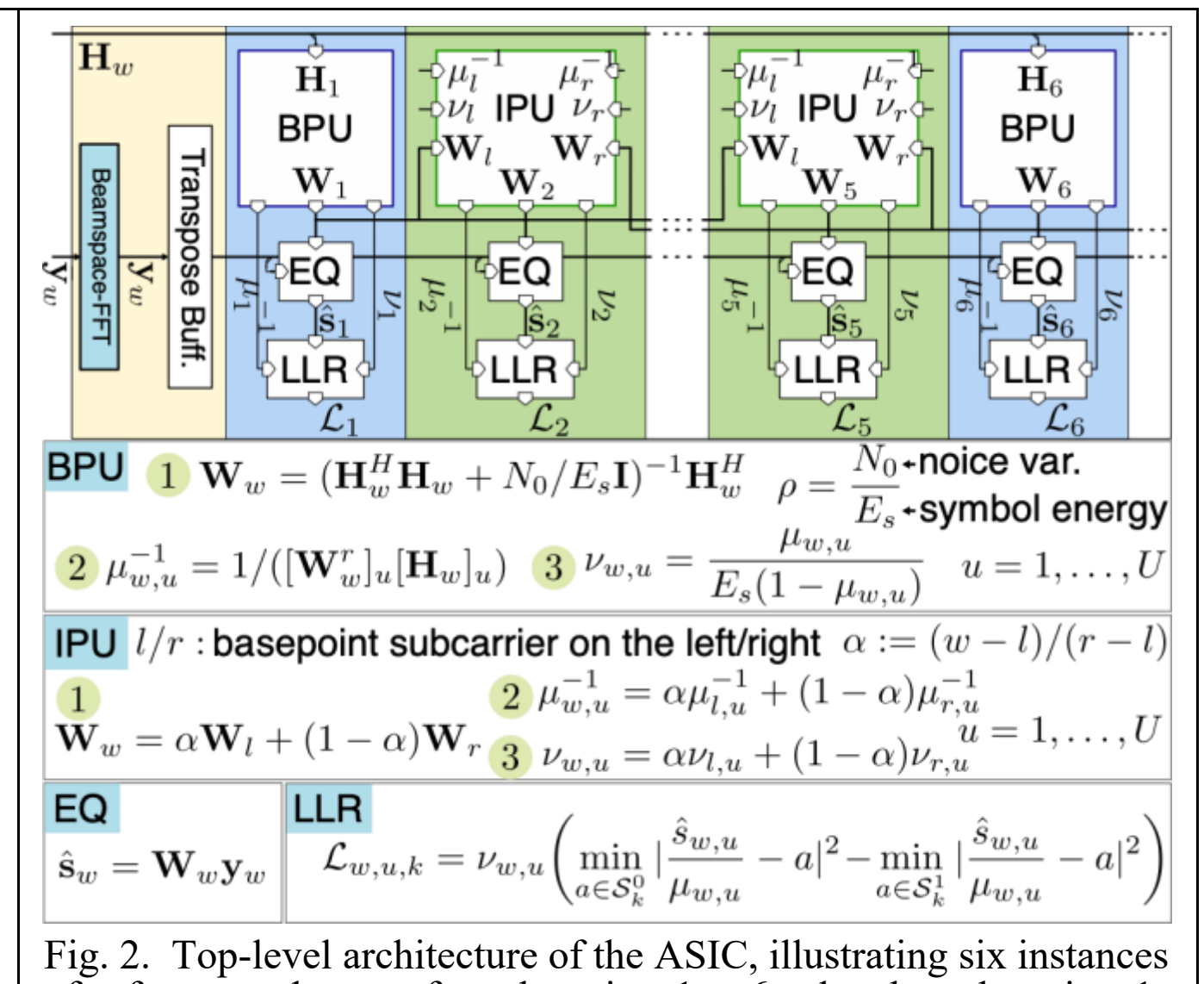

Fig. 2. Top-level architecture of the ASIC, illustrating six instances of soft-output detector for subcarriers 1 to 6, whereby subcarriers 1 and 6 are basepoints and the ones in between are interpolated. The ASIC consists of 16 parallel soft-output detectors computing LLRs.

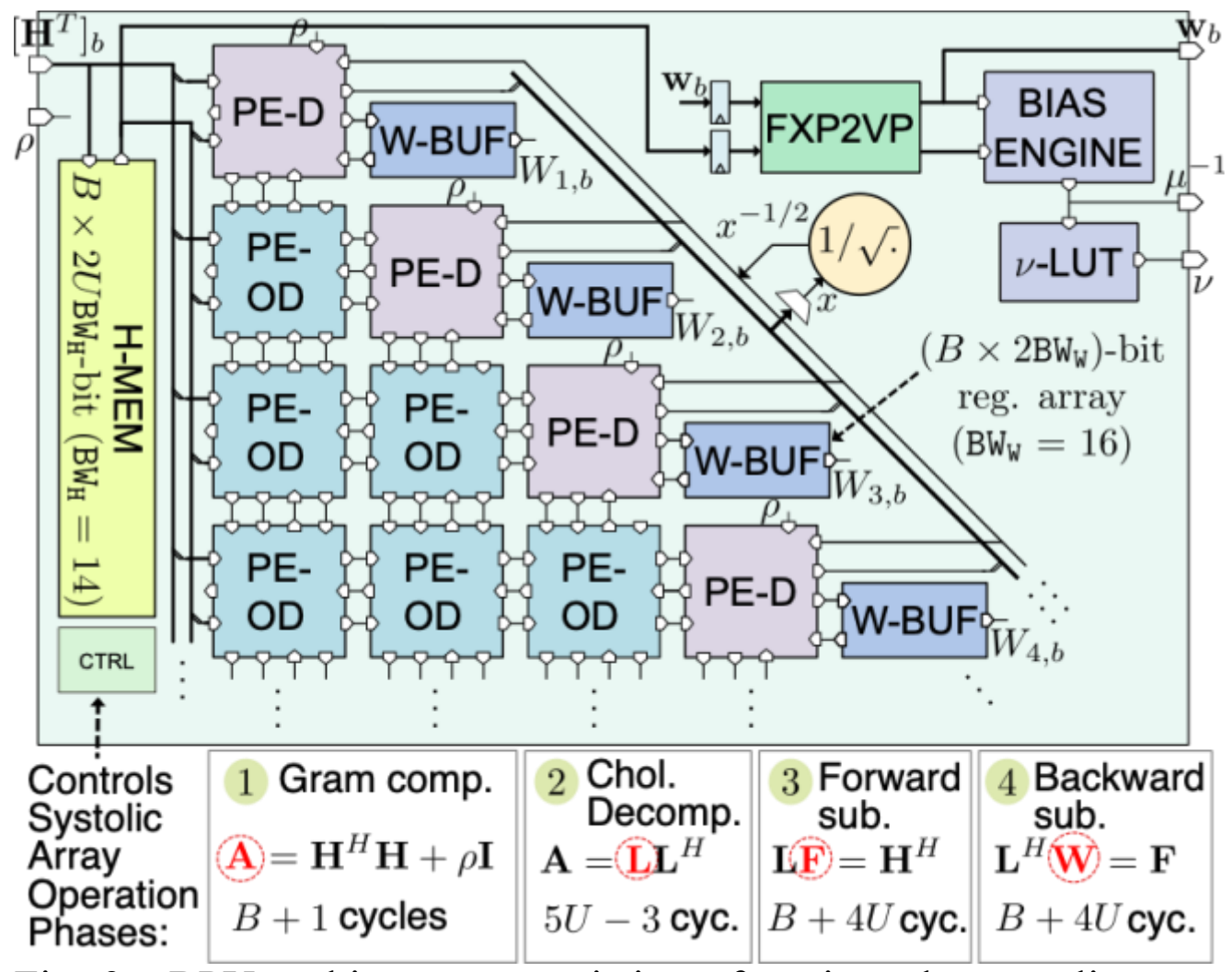

Fig. 3. BPU architecture, consisting of a triangular systolic array responsible for computing the equalization matrix $\boldsymbol{W}$, as well as the bias engine and the $\nu$-LUT, which compute the inverse bias term $\mu^{-1}$ and the post-equalization SINR term $\nu$ needed for LLR computation.

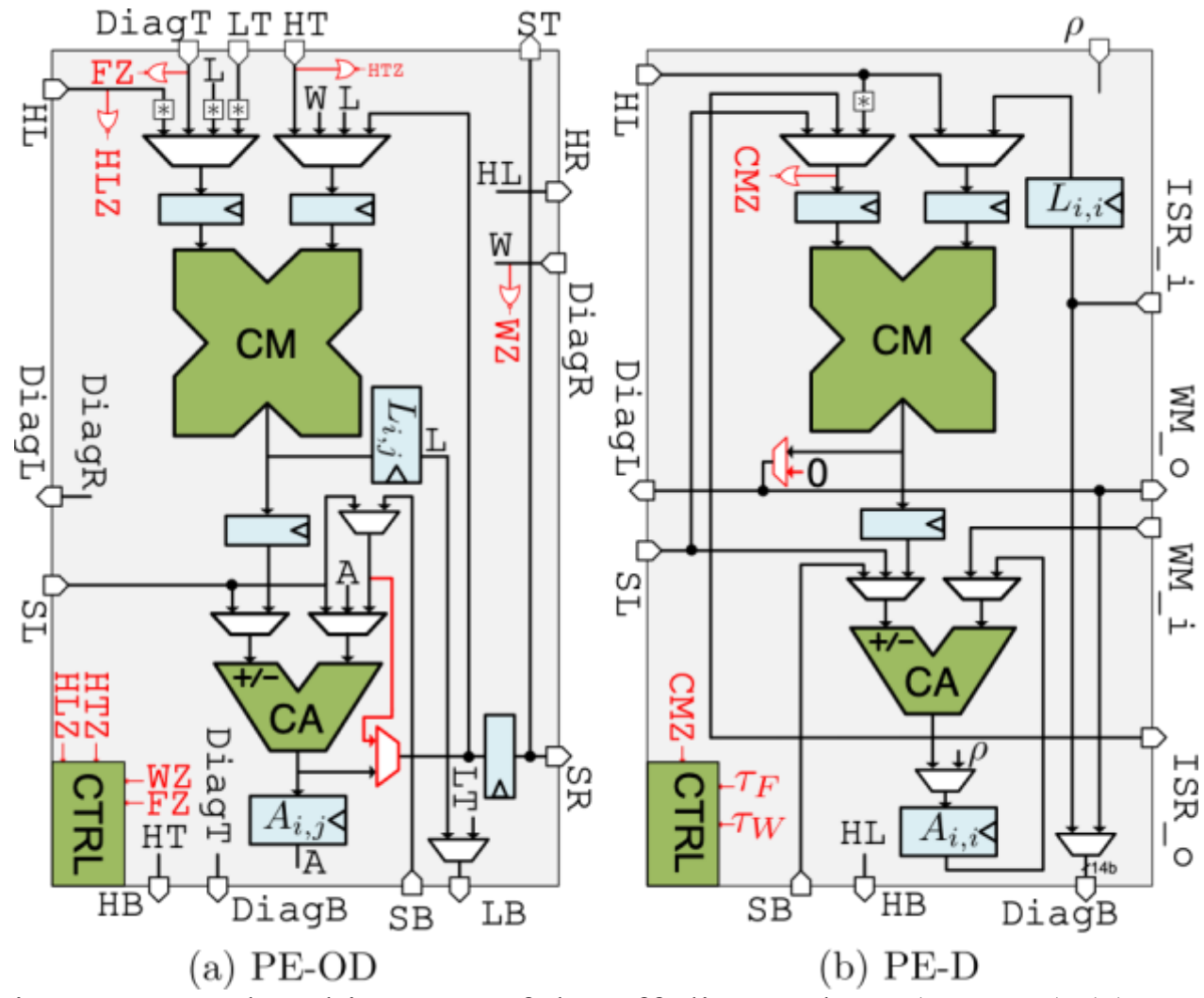

Fig. 4. Internal architecture of the off-diagonal PE (PE-OD) (a), and the diagonal PE (PE-D) (b). The PEs are designed to be able to perform the four operations shown in Fig. 3. The red components are used to perform sparsity-based signal silencing to reduce power.

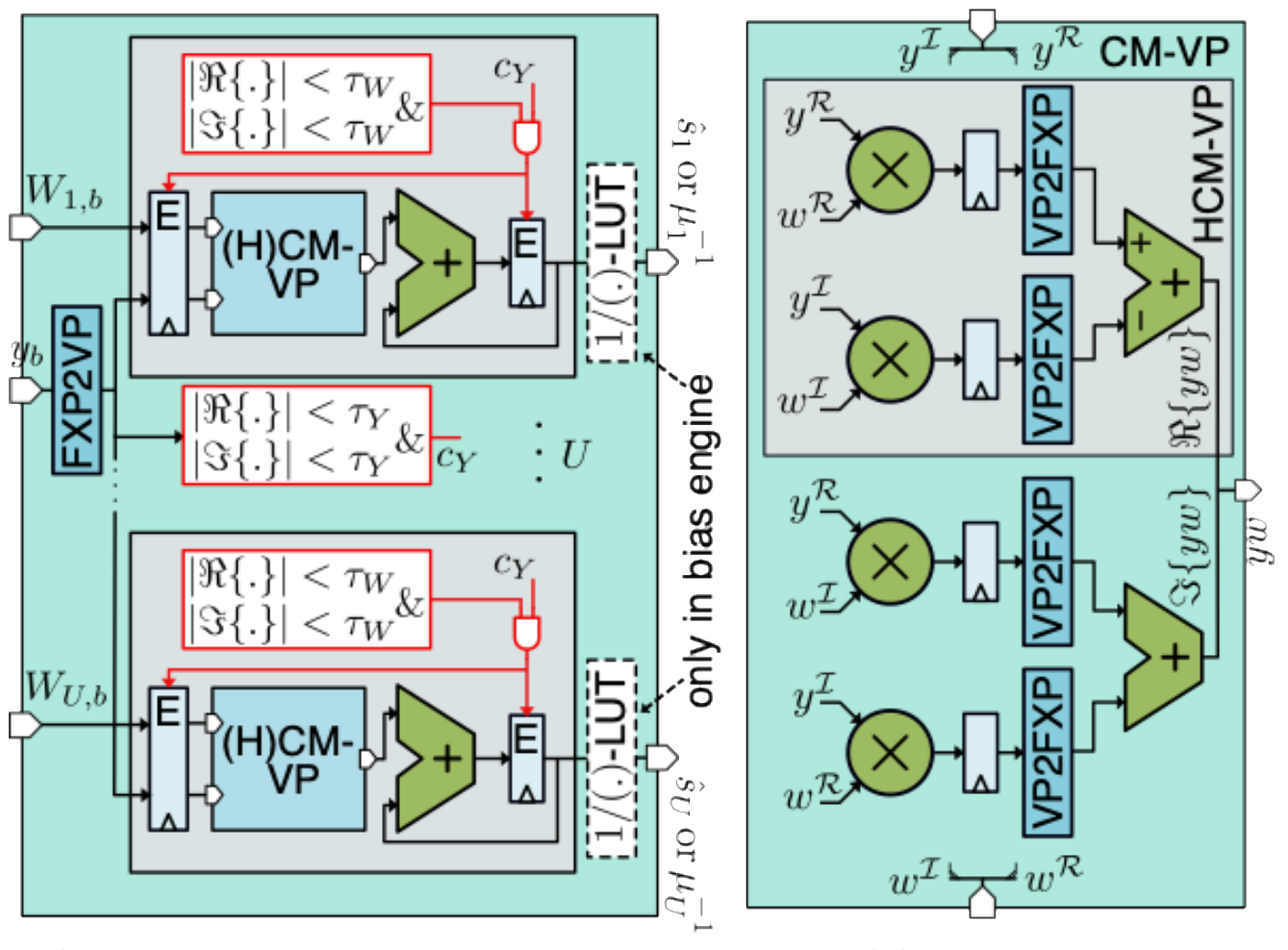

Fig. 5. Architecture of the EQ/bias engine (a) and the complex multiplier (CM) with VP inputs (b). The bias engine uses half CM-VP, while the EQ uses CM-VP. The 1/(.) LUT exist only in the bias engine. The red components are used for power saving.

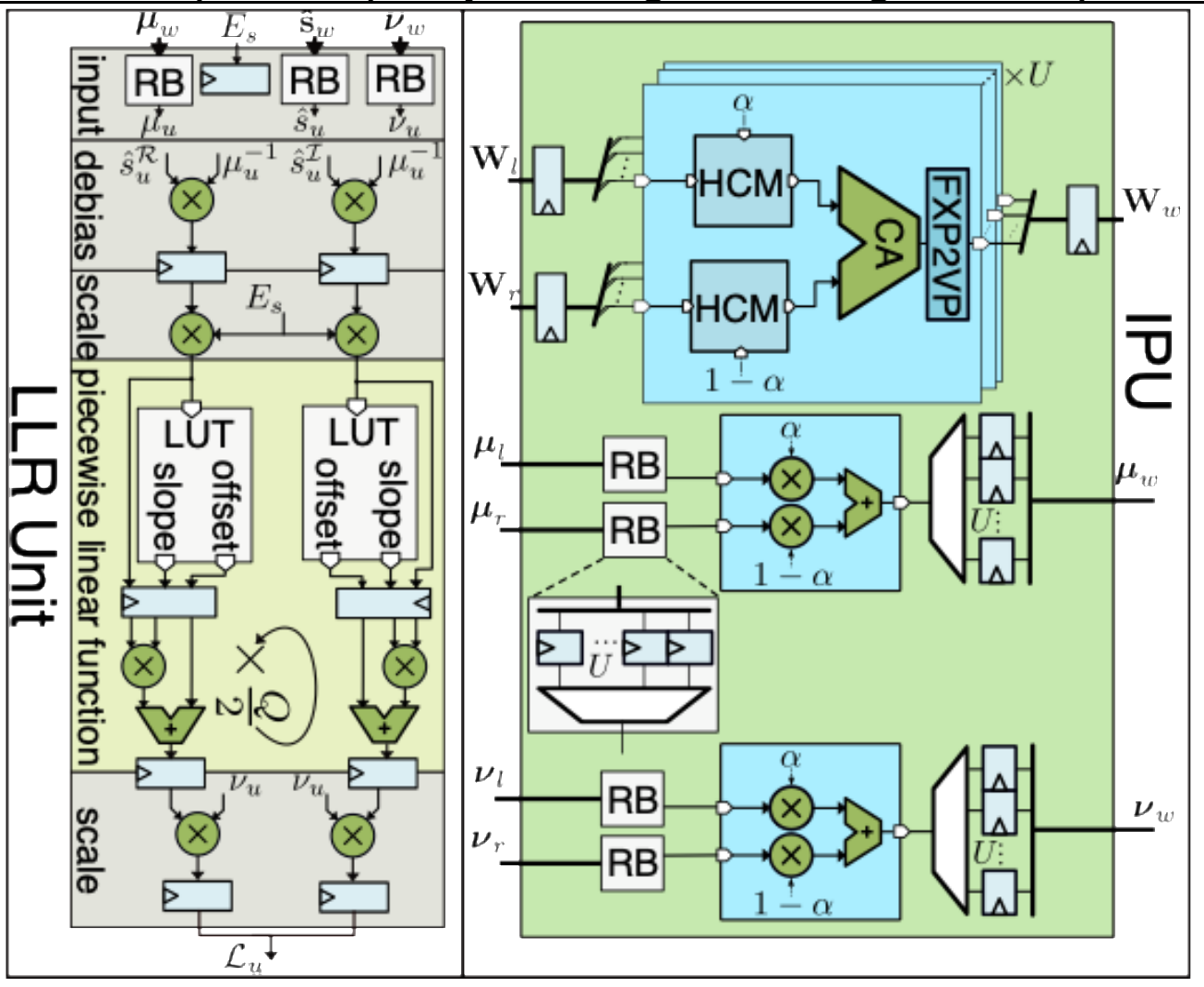

Fig. 6. Architectures of LLR engine (left), and IPU (right). The LLR unit implements a piecewise linear function iteratively for each bit $q = 1, \ldots, Q$. The IPU consists of multiply-add operations with constants determined by the interpolated subcarrier index.

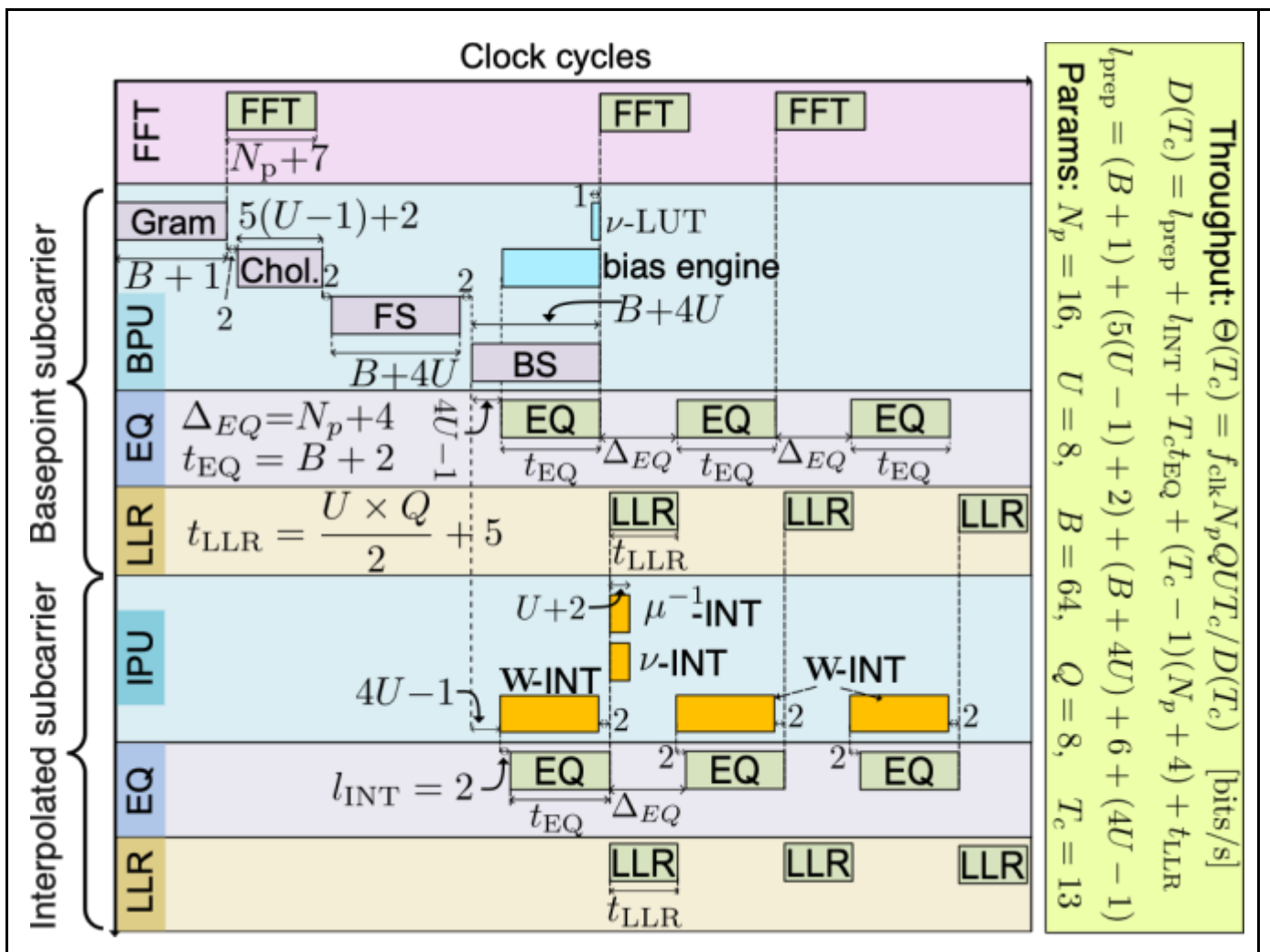

Fig. 7. Timing of modules for a basepoint subcarrier and an interpolated subcarrier (left); the throughput formula as a function of number of transmissions ($T_c$) per preprocessing operation (right). To determine the throughput, we use $T_c = 13$, as a 5G frame typically consists of one channel sounding symbol and 13 data symbols.

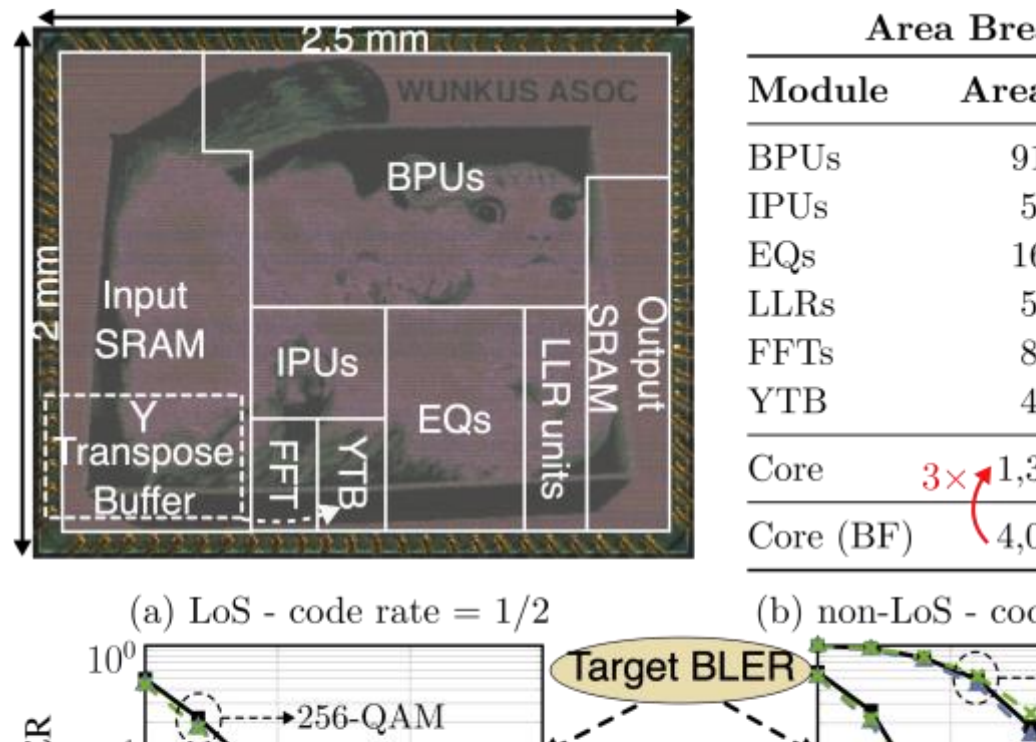

**Area Breakdown**

| Module | Area (mm²) | % |
|---|---|---|
| BPUs | 919,672 | 69.1 |
| IPUs | 54,061 | 4.1 |
| EQs | 163,077 | 12.3 |
| LLRs | 57,714 | 4.3 |
| FFTs | 81,741 | 6.1 |
| YTB | 46,975 | 3.5 |
| Core | 1,331,183 | 100 |
| Core (BF) | 4,028,195 | 304 |

3×

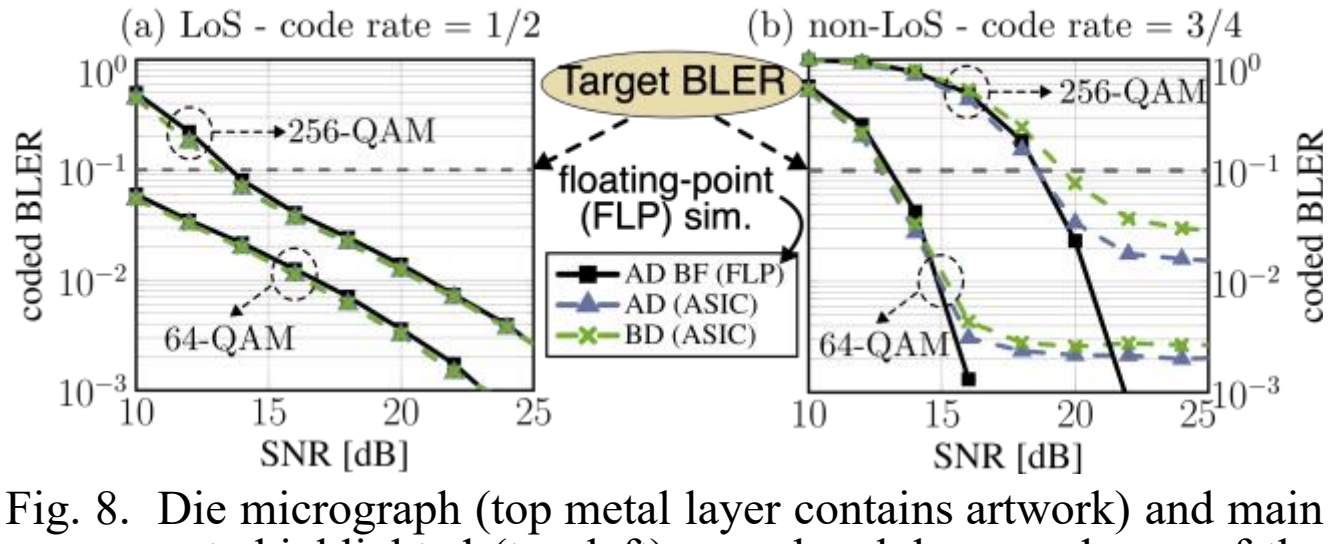

Fig. 8. Die micrograph (top metal layer contains artwork) and main components highlighted (top left); area breakdown and area of the BF baseline (top right); coded BLER of the ASIC in AD and BD mode and the BF baseline with (a) LoS and (b) non-LoS channels.

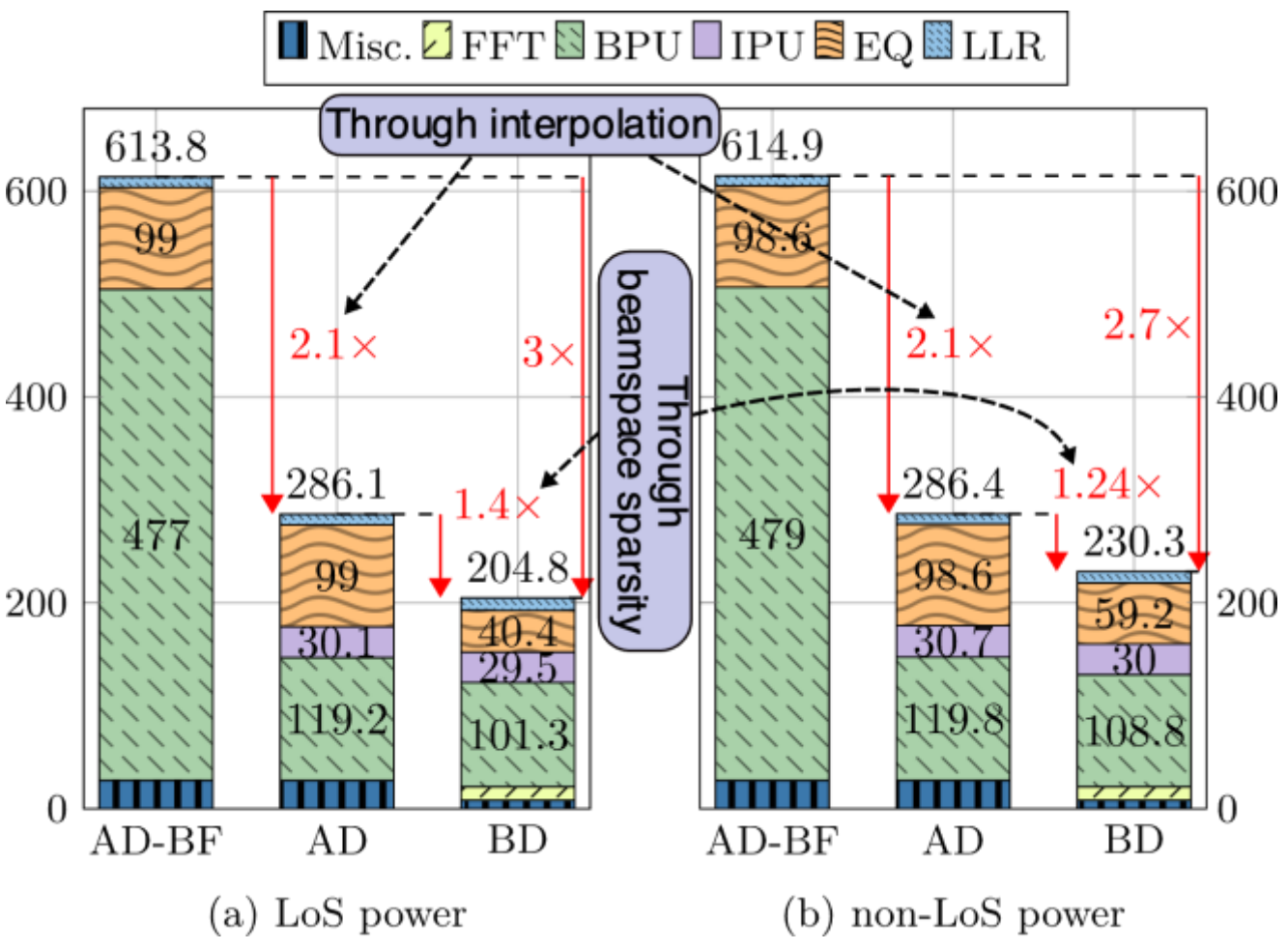

Fig. 9. Power breakdown of the chip in AD and BD mode, as well as the power of baseline BF method. The power is measured with LoS (a) and non-LoS (b) inputs at 500 MHz and 0.8V core voltage at 300K. Power of the AD-BF baseline is obtained via extrapolation.

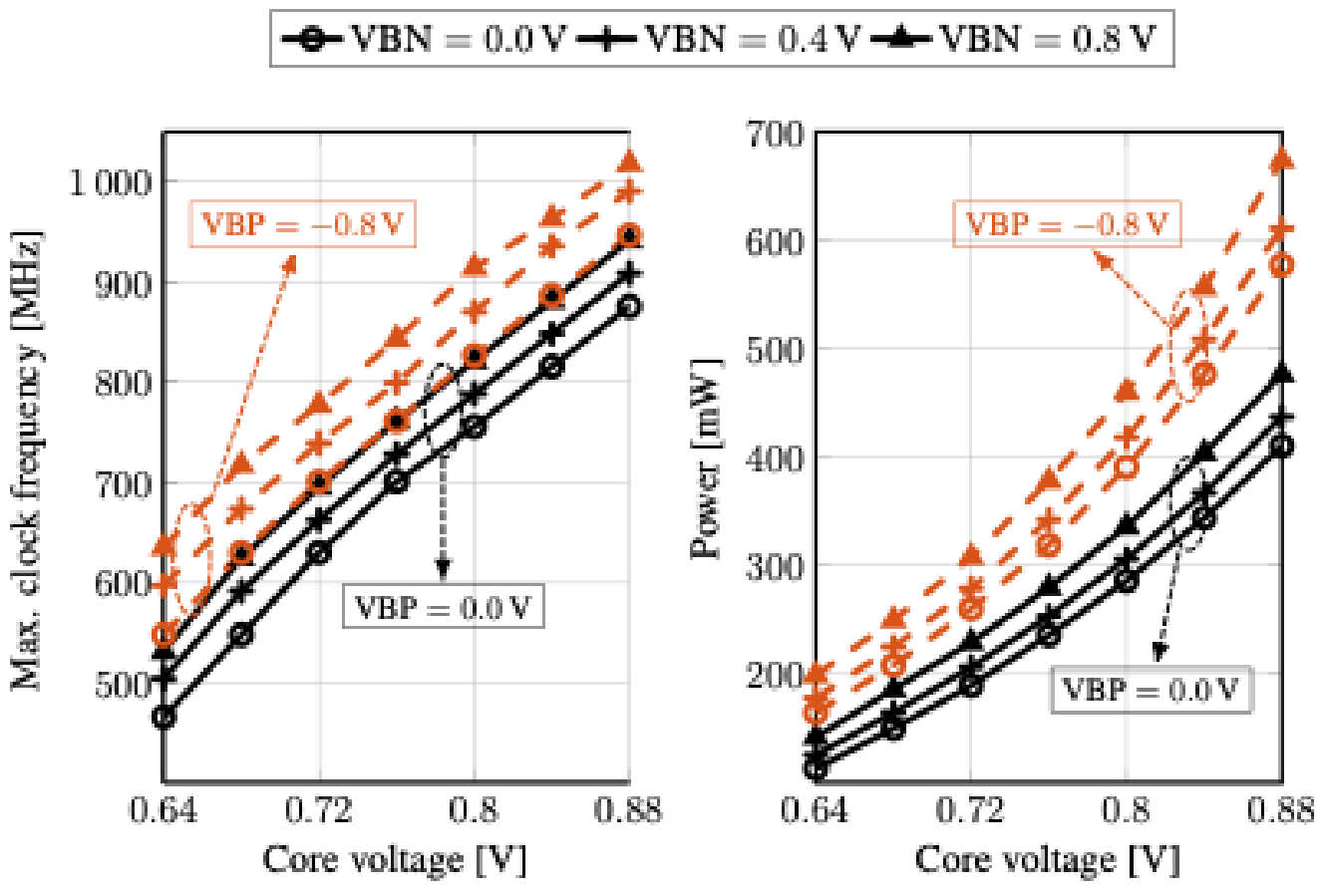

Fig. 10. Measured max. clock frequency (left) and power at the max. frequency (right) versus the applied core voltage, for a range of NMOS body bias voltages (VBN) and PMOS body bias voltages (VBP). Forward body biasing increases the throughput up to 1.4× at the expense of increased leakage power.

TABLE I. Comparison with state-of-the-art MIMO detectors.

| | This work | JSAC'25 [3] | TSP'20 [4] | JSSC'21 [5] |
|---|---|---|---|---|
| Algorithm | LMMSE | GBCD | RCG | MPD |
| MIMO dim. ($B \times U$) | 64×8 | 128×16 | 128×8 | 128×32 |
| Modulation [QAM] | 4 to 256 | 4 to 256 | 64 | 256 |
| Multi-subcarrier | **yes** | **no** | **no** | **no** |
| Preprocess. included | yes | yes | yes | no |
| Soft-outputs | yes | yes | yes | no |
| Realistic channels | yes | yes | no | no |
| Technology [nm] | 22 | 22 | 65 | 40 |
| Core volt. [V] | 0.8 | 0.8 | 1.2 | 0.9 |
| Core area [mm²] | 1.3 | 0.97 | 3.5 | 0.58 |
| Clk. freq. [MHz] | 750 | 887 | 500 | 425 |
| Throughput [Gbps] | 7.3 | 7.1 | 1.5 | 2.76 |
| Power [mW] | 286 | 787 | 557 | 220.6 |
| Energy eff. [pJ/b][ab] | 39 | 50 | 55.8 | 8 |
| Area eff. [Gbps/mm²][ab] | 5.6 | 11.5 | 11 | 114 |

(a) Technology normalized to 22 nm at 0.8 V nominal core voltage, given the assumptions: $f_{clk} \sim s$, area $\sim 1/s^2$, and power $\sim 1/(V^2)$, where $s$ is the ratio of technology nodes and $V$ is the ratio of core voltages [6]. (b) Throughput, area, and power results are scaled to a 64×8 MIMO system based on the architecture of each reference individually.